\documentclass[a4paper]{article}
\usepackage{INTERSPEECH2019}

\newcommand{\fzero}    {f_{0}}

\title{BUT VOiCES 2019 System Description}

\name{Hossein Zeinali,  
Pavel Mat\v{e}jka,
Ladislav Mo\v{s}ner, 
Old\v{r}ich Plchot, 
Anna Silnova, 
Ond\v{r}ej Novotn\'y, 
J\'an Profant,
Ond\v{r}ej Glembek, 
Luk\'a\v{s} Burget
}
\address{
Brno University of Technology, IT4I Center of Excellence, Brno, Czechia\thanks{
The work was supported by Czech Ministry of Interior project No. VI20152020025 "DRAPAK", Google Faculty Research Award program, Czech Science Foundation under project No. GJ17-23870Y, and by Czech Ministry of Education, Youth and Sports from the National Programme of Sustainability (NPU II) project "IT4Innovations excellence in science - LQ1602". It was also supported by Technology Agency of the Czech Republic project No. TJ01000208 "NOSICI",  European Union's Horizon 2020 research and innovation programme under the Marie Sklodowska-Curie grant agreement No. 748097, the Marie Sklodowska-Curie cofinanced by the South Moravian Region under grant agreement No. 665860, and by the U.S. DARPA LORELEI contract No. HR0011-15-C-0115. The views expressed are those of the authors and do not reflect the official policy or position of the Department of Defense or the U.S. Government. 
}
  }
\email{\{zeinali,matejkap,iplchot,burget,...\}@vut.cz}

\begin{document}

\maketitle
\begin{abstract}
This is a description of our effort in VOiCES 2019 Speaker Recognition challenge.
All systems in the fixed condition are based on the x-vector paradigm with different features and DNN topologies.
The single best system reaches 1.2\% EER and a fusion of 3 systems yields 1.0\% EER, which is 15\% relative improvement.
The open condition allowed us to use external data which we did for the PLDA adaptation and achieved less than ~10\% relative improvement. In the submission to open condition, we used 3 x-vector systems and also one i-vector based system.

\end{abstract}

%%%%%%%%%%%%%%%%%%%%%%%%%%%%%%%%%%%%%%%%%%%%%%%%%%%%%%%%%%%%%%%%%%%%%%%%%%%%%%%%
%%%%%%%%%%%%%%%%%%%%%%%%%%%%%%%%%%%%%%%%%%%%%%%%%%%%%%%%%%%%%%%%%%%%%%%%%%%%%%%%

%\begin{keywords}
%automatic speaker identification, deep neural networks, bottleneck features, PLDA, snorm
%\end{keywords}

%%%%%%%%%%%%%%%%%%%%%%%%%%%%%%%%%%%%%%%%%%%%%%%%%%%%%%%%%%%%%%%%%%%%%%%%%%%%%%%%
%%%%%%%%%%%%%%%%%%%%%%%%%%%%%%%%%%%%%%%%%%%%%%%%%%%%%%%%%%%%%%%%%%%%%%%%%%%%%%%%

\section{Introduction}
\label{sec:intro}
%Recent NIST SRE 2018 evaluation show the superiority of xvector based system~\cite{snyder2018x}. 
%We have done analysis between ivector and xvector on several dataset and have to confirm this - the results are in the Table~\ref{results}. 

This submission is a description of our effort in VOiCES 2019 Speaker Recognition challenge~\cite{VOiCESevalplan}.
Most of the systems are based on x-vectors~\cite{snyder2018x} with an exception of the i-vector subsystem for open condition which uses concatenation of MFCCs and Stacked bottlenecks (SBN) features~\cite{Paja_icassp2016}. Our systems utilize different features (MFCC, PLP, Mel-Filterbanks), DNN topologies and Gaussian or Heavy-tailed PLDA backend. 

Below, we present our experimental setup and the description of individual subsystems. We list the results of individual systems together with the fusion in Table~\ref{tab:results}.
%No i-vector based system made it to the fusion.

%%%%%%%%%%%%%%%%%%%%%%%%%%%%%%%%%%%%%%%%%%%%%%%%%%%%%%%%%%%%%%%%%%%%%%%%%%%%%%%%
%%%%%%%%%%%%%%%%%%%%%%%%%%%%%%%%%%%%%%%%%%%%%%%%%%%%%%%%%%%%%%%%%%%%%%%%%%%%%%%%

%%%%%%%%%%%%%%%%%%%%%%%%%%%%%%%%%%%%%%%%%%%%%%%%%%%%%%%%%%%%%%%%%%%%%%%%%%%%%%%%
%%%%%%%%%%%%%%%%%%%%%%%%%%%%%%%%%%%%%%%%%%%%%%%%%%%%%%%%%%%%%%%%%%%%%%%%%%%%%%%%
%\section{DATASET}
%Olda
\section{Experimental Setup}
\subsection{Training data, Augmentations}
\label{HXV:data}
For x-vector training we used only Voxceleb 1 and 2 dataset with 166 thousands audio files (distributed in 1.2 million speech segments) from 7146 speakers. We performed the following data augmentations based on the Kaldi recipe and created additional 5 million segments based on these augmentations:

\begin{itemize}
    \item Reverberated using RIRs\footnote{$http://www.openslr.org/resources/28/rirs\_noises.zip$}
    \item Augment with Musan\footnote{$http://www.openslr.org/17/$} noise
    \item Augment with Musan music
    \item Augment with babel noise made from Musan US-GOV speech part and Voxceleb 2 test part
\end{itemize}

%After creating a list of utterances for augmentation, a subset of 500K utterances from this list was selected and added to the training data. Afterwards, utterances with less than 500 frames and also speakers with less than 5 training utterance were removed. Finally, the training data for creating training archives contained 17054 speakers.

\subsubsection{Retransmitted NIST SRE10 close talk data}
\label{subces:retransmit_data}
In order to perform PLDA adaptation based on training data in open condition track, we made use of our dataset of retransmitted audio~\cite{JSTSP2019:Szoke}. Part of it has been benchmarked on the task of speaker verification in~\cite{Interspeech2018:Mosner}. A subset~\footnote{We used mainly telephone recordings recorded over close talk microphones} of NIST 2010 Speaker Recognition evaluations (SRE) dataset was replayed by Adam audio A7X studio monitor in numerous rooms and acoustic conditions. In each room, multiple speaker positions were considered -- sitting speaker, standing speaker and non-standard position (pointed to the ceiling, lying on the floor etc.). In addition to naturally occurring noise such as AC, vents, or common street noise coming through windows, noise source (radio receiver) was present in some sessions.

The corrupted audio was always simultaneously recorded by 31 microphones placed within the rooms. Synchronicity was governed by proprietary recording hardware.

%The original dataset consists of 932 utterances with durations of five or three minutes.
The original dataset consists of 932 utterances with 30sec durations~\footnote{The original files have duration of 5 or 3 minutes, but we take only 30 sec chunks to limit overall retransmission time.}.
There are 459 recordings from 150 female speakers and 473 recordings from 150 male speakers. The whole set was retransmitted in 5 rooms. Changes of the loudspeaker positions in some of the rooms resulted in 9 recording sessions.

\subsection{Input features}
\label{HXV:vad_features}
We use different features for several systems with this settings:

\begin{itemize}
    \item {\bf Kaldi MFCC} - Fsamp=16kHz, frequency limits 20-7600Hz, 25ms frame length, 40 filter banks, 30 coefficients + energy
    \item {\bf HTK MFCC} - Fsamp=16kHz,  frequency limits 0-8kHz, 25ms frame length, 30 filter banks, 24 coefficients + energy
    \item {\bf Kaldi PLP} - Fsamp=16kHz, frequency limits 20-7600Hz, 25ms frame length, 40 filter banks, 30 coefficients
    \item {\bf Kaldi FBank} - Fsamp=16kHz, frequency limits 20-7600Hz, 25ms frame length, 40 filter banks
    \item {\bf SBN} - Fsamp=8kHz, 80 dimensional bottleneck features trained on Fisher English, more details in Section~\ref{sec:SBN}
\end{itemize}

The Kaldi MFCC, PLP and FBank are processed with short time mean normalization over 3sec window. For HTK MFCC short time variance normalization is also applied.

\subsection{Stacked Bottleneck Features (SBN)}
\label{sec:SBN}
Bottleneck Neural-Network (BN-NN) refers to such topology of a NN, one of whose
hidden layers has significantly lower dimensionality than the surrounding
layers.  A bottleneck feature vector is generally understood as a by-product of
forwarding a primary input feature vector through the BN-NN and reading off the
vector of values at the bottleneck layer.  We have used a cascade of two such
NNs for our experiments.  The output of the first network is \emph{stacked} in
time, defining context-dependent input features for the second NN, hence the
term Stacked Bottleneck Features.

The NN input features are 24 log Mel-scale filter bank outputs augmented with
fundamental frequency features from 4 different $\fzero$ estimators (Kaldi,
Snack\footnote{http://kaldi.sourceforge.net, www.speech.kth.se/snack/}, and two
other according to ~\cite{Laskowski:LREC:2010}~and~\cite{talkin:pitch:95}).
Together, we have 13 $\fzero$ related features, see~\cite{Karafiat:IS2014} for
more details.  The conversation-side based mean subtraction is applied on the
whole feature vector. 11 frames of log filter bank outputs and fundamental
frequency features are stacked together. Hamming window followed by DCT
consisting of 0$^{th}$ to 5$^{th}$ base are applied on the time trajectory of
each parameter resulting in $(24+13)\times 6 = 222$ coefficients on the first
stage NN input.

The configuration for the first NN is $222\times D_{H}\times D_{H}\times
D_{BN}\times D_{H}\times K$, where $K$ is the number of targets. The
dimensionality of the bottleneck layer, $D_{BN}$ was fixed to 80. This was
shown as optimal in~\cite{Matejka:Odyssey2014}. The dimensionality of other
hidden layers was set to 1500.  The bottleneck outputs from the first NN are
sampled at times $t{-}10$, $t{-}5$, $t$, $t{+}5$ and $t{+}10$, where $t$ is the
index of the current frame. The resulting 400-dimensional features are input to
the second stage NN with the same topology as first stage. The 80 bottleneck
outputs from the second NN (referred as SBN) are taken as features for the
conventional GMM/UBM i-vector based SID system.

We used 8kHz SBN trained on Fisher English.

\subsection{Voice Activity Detection}
We used 2 VAD approaches:

{\bf VAD-Energy} is energy based VAD from Kaldi SRE16 recipe without any modification. Note that for FBank and PLP the Kaldi VADs from MFCC were used.

{\bf VAD-NN} consists of two carefully designed parts: a neural network (NN) which produces per-frame scores, and a post-processing stage which builds the segments based on the scores.
 
The NN was trained on the Fisher English. The input dimension is 288, while there are 2 hidden layers, each of 400 sigmoid neurons, and the final softmax layer has 2 outputs, corresponding to the classes: speech, non-speech. The NN has 277k parameters. 
 
The input features for the NN consist of 15 log-Mel filter-bank outputs and 3 Kaldi-pitch features~\cite{Ghahremani:ICASSP2014}. We apply per-speaker mean normalization estimated on the whole unsegmented recordings. Then we apply frame splicing with 31 frame-long context, where the temporal trajectory of each feature is scaled by a Hamming window and reduced to 16 dimensions by Discrete Cosine Transform. The final 288-dimensional features are globally mean and variance normalized on the NN input.

In the post-processing, we bypass the NN output softmax function (allowing us
to interpret the outputs as log-likelihoods), then we convert the two outputs to
logit-posteriors, and then we smooth the score by averaging over consecutive 31
frames.  In the final step, the speech segments were extracted by thresholding the posterior at the value of -0.5.

\section{i-vector Systems}
\label{sec:ivector}
The system is based on gender independent i-vectors~\cite{DehakN_TASLP:2010,PLDA:kenny}.
HTK MFCC with deltas and double deltas and SBN feature vectors were extracted from recordings (SBN were downsampled to 8kHz). Final feature vector is concatenation of both as they proved to perform very well in NIST SRE~\cite{Paja_icassp2016}. This system uses VAD-NN. Universal background model (UBM) contained 2048 components and was trained on Voxceleb 1 and 2 utterances from 7,146 speakers (450 hours). We then trained 600-dimensional i-vector extractor. UBM, i-vector and PLDA were trained only with clean Voxceleb data.
 
For the purpose of probabilistic linear discriminant analysis (PLDA) training we preprocessed all training, enroll and test data by means of single-channel weighted prediction error (WPE) dereverberation~\cite{IEEETrans2010:Nakatani} to suppress effects of room acoustic conditions.

\section{x-vector Systems}
\label{sec:xvector}
%Hossein I have started to write it down - feel free to modify it as you wish
All x-vectors used VAD-Energy from Kaldi SRE16 recipe~\footnote{We did not find big impact on performance when using different VAD within x-vector paradigm and it seems Kaldi simple VAD performs good for x-vector.}.
The systems were trained in Kaldi toolkit \cite{povey2011kaldi} using SRE16 recipe with modifications described below:
\begin{itemize}
    \item Using different feature sets
    \item Training networks with 9 epochs (instead of 3). We did not see any considerable difference with 12 epochs.
    \item Using modified example generation - we used 200 frames in all training segments instead of randomizing it between 200-400 frames. We also have changed generation of the training examples so that it is not random and uses almost all available speech from all training speakers in a better way.
    \item The x-vector DNN was trained on 1.2 million speech segments from 7146 speakers plus additional 5 million segments obtained with data augmentation. We generated around 700 archives that each of them contains exactly 15 training examples from each speaker (i.e. around 107K examples in each archive).
    \item The architecture of the network for x-vector extraction is shown in Table~\ref{tab:xv_baseline} and for the BIG system it is in the Table~\ref{tab:xv_BIG}.
\end{itemize}

%- describe the example generation, number of archives and other details

%\subsubsection{Training archives}

%For creating the training archives, we used Kaldi-like archive generation for our Tensorflow implementation and therefore, for both Kaldi and Tensorflow, the same configuration was used for generating two different sets of archives. Minimum and maximum number of frames in each training example are 200 and 400 respectively. Number of repeats for each speaker is 25 and maximum number of frames per archive is 2 billion. By using this configuration, two sets of 124 archives were generated for Kaldi and Tensorflow versions.

%\subsubsection{Neural net architecture, training}

\begin{table*}[tb]
\centering
\caption{\label{tab:xv_baseline}x-vector topology proposed in \cite{snyder2019speaker}. K in the first layer is used to indicate using different features with different dimensions and N is the number of speakers.}
\vspace*{+2mm}
\begin{tabular}{c|c|c}\toprule
 \textbf{Layer} & \textbf{Layer context} & \textbf{(Input) $\times$ output}\\ \cmidrule{1-3}
frame1 & $[t-2, t-1, t, t+1, t+2]$    & (5 $\times$ K) $\times$ 512 \\
frame2 & $[t]$                        & 512 $\times$ 512 \\
frame3 & $[t-2, t, t+2]$              & (3 $\times$ 512) $\times$ 512 \\
frame4 & $[t]$                        & 512 $\times$ 512 \\
frame5 & $[t-3, t, t+3]$              & (3 $\times$ 512) $\times$ 512 \\
frame6 & $[t]$                        & 512 $\times$ 512 \\
frame7 & $[t-4, t, t+4]$              & (3 $\times$ 512 $\times$ 512 \\
frame8 & $[t]$                        & 512 $\times$ 512 \\
frame9 & $[t]$                        & 512 $\times$ 1500 \\
stats pooling & $[0, T)$              & 1500 $\times$ 3000 \\
segment1 & ${0}$                      & 3000 $\times$ 512 \\
segment2 & ${0}$                      & 512 $\times$ 512 \\
softmax  & ${0}$                      & 512 $\times$ N \\
\bottomrule
\end{tabular}
\end{table*}

\begin{table*}[tb]
\centering
\caption{\label{tab:xv_BIG}BIG NN architecture. Where K is the feature dimensionality and N is the number of speakers.}
\vspace*{+2mm}
\begin{tabular}{c|c|c}\toprule
 \textbf{Layer} & \textbf{Layer context} & \textbf{(Input) $\times$ output}\\ \cmidrule{1-3}
frame1 & $[t-2, t-1, t, t+1, t+2]$    & (5 $\times$ K) $\times$ 1024 \\
frame2 & $[t]$                        & 1024 $\times$ 1024 \\
frame3 & $[t-4, t-2, t, t+2, t+4]$    & (5 $\times$ 1024) $\times$ 1024 \\
frame4 & $[t]$                        & 1024 $\times$ 1024 \\
frame5 & $[t-3, t, t+3]$              & (3 $\times$ 1024) $\times$ 1024 \\
frame6 & $[t]$                        & 1024 $\times$ 1024 \\
frame7 & $[t-4, t, t+4]$              & (3 $\times$ 1024 $\times$ 1024 \\
frame8 & $[t]$                        & 1024 $\times$ 1024 \\
frame9 & $[t]$                        & 1024 $\times$ 2000 \\
stats pooling & $[0, T)$ & 2000 $\times$ 4000 \\
segment1 & ${0}$ & 4000 $\times$ 512 \\
segment2 & ${0}$ & 512 $\times$ 512 \\
softmax & ${0}$ & 512 $\times$ N \\
\bottomrule
\end{tabular}
\end{table*}

\section{Backend}
\subsection{Heavy-tailed PLDA}
\label{HXV:backend}
Our i-vector system used HT-PLDA backend \cite{SilnovaIS18}. It was trained on VoxCeleb 1 and 2 datasets. Training set consisted of 166 thousands audio files from 7146 speakers. Length normalization, centering, LDA, reducing dimensionality of vectors to 300, followed by another length normalization were applied to all i-vectors. All i-vectors were centered using the mean computed on training data.
We fixed the size of the speaker subspace to 200. Degrees of freedom parameter was set to infinity at the training time and to 2 at scoring time. Finally, we performed adaptive score normalization as described in Section~\ref{subsec:score_norm}.

\subsection{Gaussian PLDA}
For all x-vector based systems we trained Gaussian PLDA backend. As in the case of HT-PLDA, we used concatenated data from VoxCeleb 1 and 2 for training. In this case, we train the backend only on x-vectors extracted from the original utterances augmented with reverberation and noise. X-vectors extracted from the non-augmented files were not used for backend training. 
Centering, LDA dimensonality reduction to 250 dimensions followed by length normalization was applied to x-vectors. All data were centered using the training data mean.
Speaker and channel subspace size was set to 250 (i.e full rank).
Same as in the case of HT-PLDA, we applied adaptive score normalization described in Section~\ref{subsec:score_norm}.

\subsection{Adaptation (ADAPT)}
For open condition, we used 280k files of BUT retransmitted data (see Section~\ref{subces:retransmit_data}) to perform domain adaptation by model interpolation. That is, we train smaller G-PLDA model on retransmitted data, size of both speaker and channel subspaces was fixed to 150.  The final adapted model is derived from the two G-PLDA models so that the modeled within- and across-speaker covariance matrices are a weighted combination of the covariance matrices from the constituent models. Similarly, the model means are also interpolated. Interpolation weights are set to 0.6 for the original model and 0.4 for the adaptation one. The systems which use this adaptation are denoted ADAPT in the Table~\ref{tab:results}.

\subsection{Score normalization}
\label{subsec:score_norm}
%The goal of score normalization is to reduce within trial variability leading to improved performance, better calibration, and more reliable threshold setting. 
We used adaptive symmetric score normalization (adapt S-norm) which computes an average of normalized scores from Z-norm and T-norm~\cite{PLDA:kenny,ICSLP:Matejka}. 
In adaptive version~\cite{ICSLP:Matejka,ICASSP2005:Sturim,Odyssey2006:Zigel}, only part of the cohort is selected to compute mean and variance for normalization. Usually $X$ top scoring or most similar files are selected, where $X$ is set to be 400 for all experiments.
The cohort is created from training data and consist of approximately 15k files, (two files per speaker).

\section{Calibration \& Fusion}
\label{sec:fusion}

\begin{table*}[th]
\caption{ Development results }
\label{tab:results}
  \centerline{
    \begin{tabular}{c c l l c | c c c | c c c} 
    \toprule
    &    &    &     &     & \multicolumn{3}{c}{VOiCES dev} & \multicolumn{3}{|c}{SITW core-core}\\
    && system & VAD & FEA & MinDCF & PRBEP & EER & MinDCF & PRBEP & EER \\
    \midrule \midrule
    & 1 & x-vector & Kaldi & FBANK  & 0.141 & 1908.8 &  1.23 &    0.188 & 461.4 & 1.80\\  %FBANK_REV_9EP
    fixed & 2 & x-vector & Kaldi & PLP  & 0.163 & 2204.3 &  1.44 &    0.191 & 464.6 & 1.92\\    %PLP_REV_NOI_BAB
    & 3 & x-vector BIG & Kaldi & MFCC & 0.163 & 2186.8 & 1.29 & 0.177 & 430.2 & 1.77 \\   %MFCC_BIG
    \midrule
    &4 & i-vector & VAD-NN & MFCC+SBN & 0.428 & 5911.6 & 4.46 & 0.275 & 693.4 & 3.19 \\   %IVECS_MFCC_BN_HTPLDA
    open &5 & x-vector ADAPT & Kaldi & FBANK & 0.146 & 1954.0 & 1.13 & 0.202 & 495.8 & 1.99 \\   %FBANK_REV_ADAPT_OPEN
    &6 & x-vector ADAPT & Kaldi & PLP & 0.157 & 2123.3 & 1.31 & 0.195 & 481.3 & 2.11 \\   %PLP_REV_ADAPT_OPEN
    \midrule
    fixed & 1+2+3 & PRIMARY 1 & & & 0.122 & 1647.1 & 1.04 &    0.17 & 427.3 & 1.65 \\ %FBANK_REV_9EP+PLP_REV_NOI_BAB+MFCC_BIG
    fixed & 1 & CONTRASTIVE 2 & & & 0.141 & 1908.8 &  1.23 &    0.188 & 461.4 & 1.80 \\  %FBANK_REV_9EP  
    open & 3+4+5+6 & PRIMARY 1 & & & 0.119 & 1596.1 & 1.00 &    0.17 & 432.1 & 1.73 \\    %IVECS_MFCC_BN_HTPLDA+MFCC_BIG+PLP_REV_ADAPT_OPEN+FBANK_REV_ADAPT_OPEN
    \bottomrule  
   \end{tabular}
   %}
  }
\end{table*}

The submission strategy was one common fusion trained on the labeled VoiCES development data~\cite{VOiCEScorpus,VOiCESevalplan}. Each system provided log-likelihood ratio scores that could be subjected to score normalization. These scores were first pre-calibrated and then passed into the fusion. The output of the fusion was then again re-calibrated. 

Both calibration and fusion were trained with logistic regression optimizing the cross-entropy between the hypothesized and true labels on a development set. Our objective was to improve the error rate on the development set itself, but we were also monitoring error-rate trends on Speakers In The Wild dataset. Results of individual systems and fusions are listed in Table~\ref{tab:results}.

\bibliographystyle{IEEEbib}

\bibliography{biblio}

\end{document}